\documentclass[twocolumn,aps,prl,floats]{revtex4}
\usepackage{graphics}
\usepackage{amssymb,amsmath}

\begin{document}

\title{ Environment effects
on effective magnetic exchange integrals and local spectroscopy of extended 
strongly correlated systems.}
\author{Marie-Bernadette LEPETIT} 
\affiliation{Laboratoire de Physique Quantique, IRSAMC~/~UMR~5626,
Universit\'e Paul Sabatier, 118 route de Narbonne, F-31062 Toulouse
Cedex 4, FRANCE}
\author{Nicolas SUAUD}
\affiliation{}
\author{Alain GELLE}
\affiliation{Laboratoire de Physique Quantique, IRSAMC~/~UMR~5626,
Universit\'e Paul Sabatier, 118 route de Narbonne, F-31062 Toulouse
Cedex 4, FRANCE}
\author{Vincent ROBERT}
\affiliation{Laboratoire de Physique Quantique, IRSAMC~/~UMR~5626,
Universit\'e Paul Sabatier, 118 route de Narbonne, F-31062 Toulouse
Cedex 4, FRANCE}
\altaffiliation[Permanent address: ]{Institut de Recherches sur la Catalyse, Laboratoire de Chimie Théorique, Bd. du 11 novembre 1918, 69619 Villeurbanne Cedex, FRANCE}

\date{Accepted for publication in the Journal of Chemical Physics on the $5^{th}$ of november 2002}

\begin{abstract}
The present work analyzes the importance of the different components
of the environment effects on the local spectroscopy of extended
strongly correlated systems. It has been found that the usual formal
charge definition of the charge transfer and Madelung potential are
far too crude for an accurate determination of the local excitation
energies in embedded fragment calculations. A criterion for the
validation of the embedding against the infinite system density of
states has been proposed.  \pacs{}
\end{abstract}

\maketitle

\section{Introduction}
\label{sec:Intro}
In the last twenty years the science of strongly correlated materials
has attracted a lot of attention. Indeed, chemists have synthesized a
large number of new families of materials which present unusual and
fascinating properties directly related to the strongly correlated
character of their electronic structure (to cite only a few systems,
one can think of the copper oxides presenting $d$-wave, high-$T_c$
super-conductivity~\cite{htc}, manganites with giant
magneto-resistance~\cite{mang}, Prussian Blue analogs with
photo-magnetic properties~\cite{phmg}, etc.).  In all these materials
the electrons responsible for their spectacular properties are few
(per unit cell), usually unpaired and localized both spatially and
energetically near the Fermi level. That is to say that the wave
function of these systems is essentially multi-configurational and
cannot be correctly treated by single-reference based methods such as
Hartree-Fock plus perturbation theory or even Density
Functional~\cite{dft}, the only {\em ab initio} methods to be tractable on
infinite crystals.  Physicists have thus relied on semi-empirical
models such as the Heisenberg~\cite{heis}, $t-J$~\cite{t-J} or
Hubbard~\cite{hubbard} Hamiltonians in order to describe these
strongly correlated materials.  These effective models are of
valence-bond (VB) type and belong to the zero differential overlap
(ZDO) class. They are aimed at describing the effective microscopic
interactions between the set of valence electrons responsible for the
macroscopic low energy properties.  For instance, this is in this
respect that the famous $t-J$ model has first been derived by Zhang
and Rice {\it et al}~\cite{t-J} as a model for the high-$T_c$
copper oxides. Indeed, the $t-J$ model describes the dominant
interactions between the magnetic electrons of the $C\!uO_2$
superconducting planes. These planes can be seen as a $C\!u^{2+}$
square lattice, each copper ion supporting (before doping) one
$d_{x^2-y^2}$ magnetic electron.  These unpaired, Fermi level
electrons (considered to be the important ones for the
superconductivity properties) only interact between each others
through super-exchange processes (via the bridging oxygen anions
$O^{2-}$) and through effective hopping integrals (when the planes are
doped as in the super-conducing phases), i.e. through a $t-J$ type of
Hamiltonian.

One sees immediately that such representations necessitate i) to
clearly identify the pertinent electrons and orbitals responsible for
the macroscopic low energy properties ---~they will be called from now
on active electrons and active orbitals~--- ii) to determine the
dominant interactions between these electrons iii) finally to evaluate
their relative amplitudes. While the nature of the active electrons
and orbitals, and even the form of the dominant interactions, are most
of the time relatively easy to derived using simple chemical
considerations, the quantitative determination of the interactions
amplitudes is a much more troublesome problem, even-though as
crucial. Indeed, the low energy properties of the macroscopic systems
are highly dependent on the relative amplitudes of the different
effective integrals, since the latter can govern properties as
important as the metallic versus insulating character of a compound.
Unfortunately, experimental data are, most of the time, unable to
totally determine them and it is necessary to rely on {\em ab initio}
quantum-chemical calculations for this purpose. Since these effective
interactions (hopping,  magnetic exchange, on-site bi-electronic
repulsion, etc.) are essentially local (see
references~\cite{revue,loca} for detailed analyses of the locality of
the different effective interactions) and involve a small number of
bodies (most of the time only one or two), they can be accurately
determined from the computation of the local excitation energies and
related states. For instance, in the high-$T_c$ copper oxides, the
magnetic coupling  between two active $d_{x^2-y^2}$ electrons can
be determined as the local singlet to triplet excitation energy
between two nearest neighbors, oxygen bridged, copper atoms.  The
embedded cluster technique, where the local valence excitations are
computed on an adequate fragment of the crystal, has proved to be
one of the most efficient technique for this purpose. Indeed, the full
power of the {\em ab initio} molecular-spectroscopy methods can be used on
such finite systems. One can cite the successes obtained on high-$T_c$
copper oxides~\cite{cuo}, vanadates oxides~\cite{vana,vana3},
fluoroperovskites~\cite{perov}, etc. For instance, the
magnetic exchange in the high-$T_c$ superconductor
$N\!d_2C\!uO_4$ has been computed within experimental accuracy
($J_{calc.}=-126.4\,meV$ being the computed value in
reference~\cite{cuo} and $J_{exp.}=-126 \pm 5\,meV$ being the measured
experimental value~\cite{Nd}).

As we will see in details later on, the local character of the
effective interactions refers to the local character of the electronic
processes involved in the local excitations, once the mono-electronic
part of the wave function is defined, that is, once the shape of the
fragment orbitals and their energies are defined.  It is however
obvious that the latter, and in consequence the effective interactions
amplitudes, are strongly influenced by the fragment crystalline
surrounding. Indeed, the orbitals and orbital energies of an isolated
fragment are very different from the orbitals and orbital energies of
the same fragment in a crystalline environment.  It is therefore
crucial to embed the studied fragment into an environment reproducing
the main effects of the rest of the crystal.  

The purpose of this
paper is to analyze, review and clarify the requirements for a good
embedding and in this light to discuss the most common practices. We
will also propose a general technique for checking the quality of an
embedding and study the sensitivity of some effective parameters, such
as the magnetic super-exchange and the effective hopping integrals, to
the quality of the embedding and the crystalline environment.  The
next section will therefore be devoted to the analysis of the
environment-fragment interactions and the proposal of a simple
criterion measuring the embedding quality. Section three will review
the usual practices, exemplify the influence of the crystalline
environment on real systems and point out the larger importance of the
different embedding components on the effective parameters than is
commonly assumed. Finally, the last section will be devoted to 
conclusions and perspectives.

\section{Interactions between the fragment and the rest of the crystal}
 
The interactions between a selected fragment and the rest of the
crystal can be classified into two different types~: the effects on
the fragment of the crystal average electrostatic field and the
effects of its fluctuations.

Let us first start with the second kind~: the electrostatic and spin
fields fluctuations, that is the instantaneous response of the rest of
the crystal to a particular electronic configuration of the
fragment. In other words we are talking about the dynamical
correlation effects of the rest of the crystal on the fragment valence
states under consideration. The spatial separation between the active
centers and their crystalline environment allows us to conduct a
multi-polar analysis of the different contributions to the excitation
energies, following the same pattern used in the description of the
inter-molecular interactions.  Along this line, it is very simple to
see that the leading dynamical-correlation contributions are the
polarization responses of the environment first to a local charge,
then to a dipole moment or transition dipole moment on the
fragment~\cite{revue}.  In other words, these effects correspond to
excitations from the average environment (chosen as its zeroth order
description) in response to a particular fragment configuration. In a
perturbative scheme they can be described as the coupling between
single excitations in the environment to a local charge, a local
dipole or a local transition dipole on the fragment.  From multi-polar
expansion combined to second order perturbation theory, it is easy to
see that the leading contributions come from the interactions between
a charge on the fragment and a dipole on the environment. Such effects
decrease as $(1/R^2)^2$, that is quite rapidly. For a complete
description and analysis of the different contributions one can refer
to~\cite{revue}. Provided that the largest $1/R^4$ terms ---~i.e. the
terms coming from single-excitations on the first neighbors of the
magnetic centers~--- are correctly treated, the rapid decrease of
these effects as a function of $R$ allows us to consider the rest of
them as reasonably negligible.  Indeed, while the first neighbor shell
of atoms surrounding the magnetic sites are usually at a distance of
$2$\AA~to $3$\AA (from $3.8$a.u. to $5.7$a.u.) the second
neighbor distances are usually of the order of $3$\AA~to $5$\AA, 
($5.7$a.u. to $9.5$a.u.). The second neighbor contributions can
therefore be expected to be of an order of magnitude $10^{-3}$ to
$10^{-4}$ smaller than the on-site polarization contributions. One can
therefore consider that including in the studied fragment, the first
coordination shell of all the atoms involved in the local excitations
under consideration, is enough to take into account the essential
contributions to the modification of the effective interactions by the
environment dynamical correlation effects.

Let us now examine the effects on the fragment, of the average
electrostatic field of the rest of the crystal. At this point it is
important noticing that a majority of the strongly correlated
materials are ionic or at least partly ionic. This is for instance the
case of the high-$T_c$ superconducting cuprates for which the
super-conducting planes are hole-doped $C\!u^{2+}O^{2-}_2$
lattices. Similar observations can be made for vanadate, manganites,
cobaltites and even the organic conductors which are charge transfer
salts. It is therefore clear that electrostatic effects such as the
Madelung potential have to be properly treated and that, unlike the
field fluctuations, the average electrostatic and spin fields of the
environment strongly affect the fragment under consideration. This
problem has been carefully studied by Barandiar\'an and Seijo in the
case where the fragment borders do not cut any covalent bond. They
analyzed the crystalline environment contributions as Coulomb
interactions, quantum exchange interactions and orthogonality of the
fragment orbitals to the orbitals of the rest of the crystal (quantum
orthogonality interactions)~\cite{pszoila}. Out of these one can
separate the short range contributions~: short range Coulomb
repulsions, quantum exchange interactions and quantum orthogonality,
from long range ones, that is essentially Coulomb interactions such as
the Madelung potential. The latter can be easily reproduced using the
dominant terms of a multi-polar expansion, that is for an ionic (or
even partly ionic) crystal the point charge Madelung potential as
described using the Ewald~\cite{ewald} formulation or, as it is more
commonly done, by a large enough set of point charges modified at the
set borders by the Evjen's procedure~\cite{evjen}. The short range
contributions necessitate a more sophisticated treatment since the
fragment orbitals should be orthogonal to the orbitals (non-included
in the calculation) supporting the electrons of the rest of the
crystal, that is essentially with the electrons located on the
fragment first-shell neighboring atoms. The resulting exclusion of
regions in the $\mathbb{R}^3$ space for the fragment electrons
wave-function has proved to be well treated by total ion
pseudo-potentials (TIP) approaches. Indeed, the TIPs are explicitly
derived in order to render on the fragment orbitals the quantum
orthogonality, quantum exchange and short range Coulomb potential
effects~\cite{Pitzer,pszoila}.  As for the Madelung potentials, the
short range contributions acts essentially at the mono-electronic
description level (Hartree-Fock).  Indeed, both the electrostatic
potential and the TIPs modify only the mono-electronic integrals. Thus
these effects affect the Fock operator and strongly influence the
shape of the fragment orbitals and their energies.

While the determination of the TIPs modeling the short range
contributions is quite unambiguous, the determination of the Madelung
field is more tricky. Even-though experimental determinations of the
electrostatic potential at points locations may be available, this is
usually not the case and the question of what are the average charges
supported by the different atoms remains open. Most authors use simple
chemical considerations such as the standard atomic oxidation states
and charge equilibrium to derive the formal charges supported by the
different atoms in the crystal. For instance, $N\!d^{3+}$, $C\!u^{2+}$
and $O^{2-}$ in the neodymium-doped superconductor $N\!d_2C\!uO_4$, or
$N\!a^{1+}$, $V^{4.5+}$, $O^{2-}$ in the famous inorganic spin-Peierls
$\alpha^{\prime}N\!aV_2O_5$ compound. Such simple considerations
assume totally ionic systems. However in most systems the ionicity is
not as strictly total and for systems presenting iono-covalent bonds
the field created by the formal charges does not accurately reproduce
the real Madelung potential. The problem of determining the degree of
polarization of such systems or the real charges supported by the
atoms is therefore crucial.  That was for instance the case in the
$\alpha^{\prime}N\!aV_2O_5$ compound which was commonly assumed in the
literature to be $N\!a^{1+}V_2^{4.5+}O_5^{2-}$, while crystal
{\em ab initio} mean-field calculations (Hartree-Fock calculations with
valence triple zeta + polarization type of basis sets~\cite{bs} run
using the CRYSTAL code~\cite{crystal}) yield M\"ulliken atomic charges
of $N\!a^{0.95+}V_2^{2.3+}O_2^{0.95-}O_2^{1.3-}O^{1.15-}$. It is well
known that the M\"ulliken charges are strongly dependent of the chosen
basis set and in particular of its spatial extension. However, the
above mean-field calculations clearly exhibit a strongly-covalent
triple bond between each vanadium atom and its corresponding apical
oxygen~\cite{vana}, a strong donation from the apical oxygen to the
vanadium atom ---~in order to form the third bond~--- and a weaker donation
of the other oxygen anions to this $V\equiv O$ system. This result,
which couldn't have been derived from simple chemical considerations,
have been latter confirmed by embedded fragments Complete Active Space
Self Consistent Field (CASSCF) calculations~\cite{vana}.  As a
conclusion, the choice of the atomic charges generating a physically
correct Madelung field is a crucial issue that cannot be solved by
simple charge transfer considerations.

The problem can be solved by going back to the question of how the
electrostatic potential acts on the fragment wave-function. As already
mentioned, the potential generated by the rest of the crystal acts
through the mono-electronic integrals, and thus, on the orbital shapes
and orbital energies through the Fock operator definition. All these
effects belong to zeroth-order description of the fragment and are
present at the Hartree-Fock level. It is therefore possible to
calibrate these effects by a direct comparison of the Hartree-Fock
description of the embedded fragment and the Hartree-Fock description
of the infinite crystal. Indeed, the density of states (DOS) of the
infinite crystal ($\rho^{crys}({\cal E})$) can be projected on the
atomic orbitals involved in the fragment under consideration and
compared with the embedded fragment orbitals atomic contents and 
energies. It is easy to define a projected density of states (PDOS)
function for the finite fragment in a similar way as in an infinite
system. The major difference between the two is that the finite system
PDOS is a discrete function. If $\rho^{crys}_\mu({\cal E})$ is the
PDOS of the entire crystal on the atomic orbital $\mu$,
\begin{eqnarray} \label{eq:cdos} \hspace*{-0.35cm}
\rho^{crys}_\mu({\cal E}) &=& {1 \over V_{B\!Z}} 
\sum_{n, \nu} \int_{B\!Z}   c^*_{\nu n}(\vec k)\, S_{\nu \mu} \,
c_{\mu n}(\vec k)\,   \delta({\cal E - E}_n(\vec k)) \,d\!\vec k \hspace*{2eM}
\end{eqnarray}
the embedded fragment {\em density of states}, $\rho^{frag}_\mu({\cal
E})$, can be defined in a similar way as
\begin{eqnarray} \label{eq:fdos}
\rho^{frag}_\mu({\cal E}) &=&  \sum_j \sum_\nu f^*_{\nu j}\,S_{\nu \mu} \, 
f_{\mu j} \, \delta({\cal E - E}_j) 
\end{eqnarray}
where $B\!Z$ refers to the Brillouin zone, $V_{B\!Z}$ to its volume,
$\vec k$ is the wave vector, $c_{\mu n}(\vec k)$ is the coefficient of
the $n^{\rm th}$~-~band crystalline orbital at point $\vec k$ on the
atomic orbital $\mu$ (the star representing the complex conjugate) and
${\cal E}_n(\vec k)$ the associated orbital energy. Finally, $f_{\mu
j}$ is the usual coefficient of the $j^{\rm th}$ molecular orbital of
the fragment on the atomic orbital $\mu$ and ${\cal E}_j$ the
associated orbital energy. \\ If the embedding correctly reproduces
the average effects of the rest of the crystal on the fragment, the
discrete function $\rho^{frag}_\mu({\cal E})$ should be located at the
same energetic positions as the crystal PDOS $\rho^{crys}_\mu({\cal
E})$, and the main contributions of both functions should coincide.

\section{Some practical examples}

As already mentioned, the most usual practice used to define a
fragment embedding consists in \begin{itemize}
\item building the Madelung potential using formal charges, 
\item using TIPs on the first shell of cations in order to set up the
main exclusion effects.
\end{itemize}
In the following subsections, we will analyze on different examples 
the validity and limitations of such a procedure, as well as the
influence of the different parameters on the embedding quality.

\subsection{The importance of a covalent bond~: 
the $N\!aV_2O_5$ compound} 

\begin{figure}[h]
\centerline{
\resizebox{4.0cm}{3cm}{\includegraphics{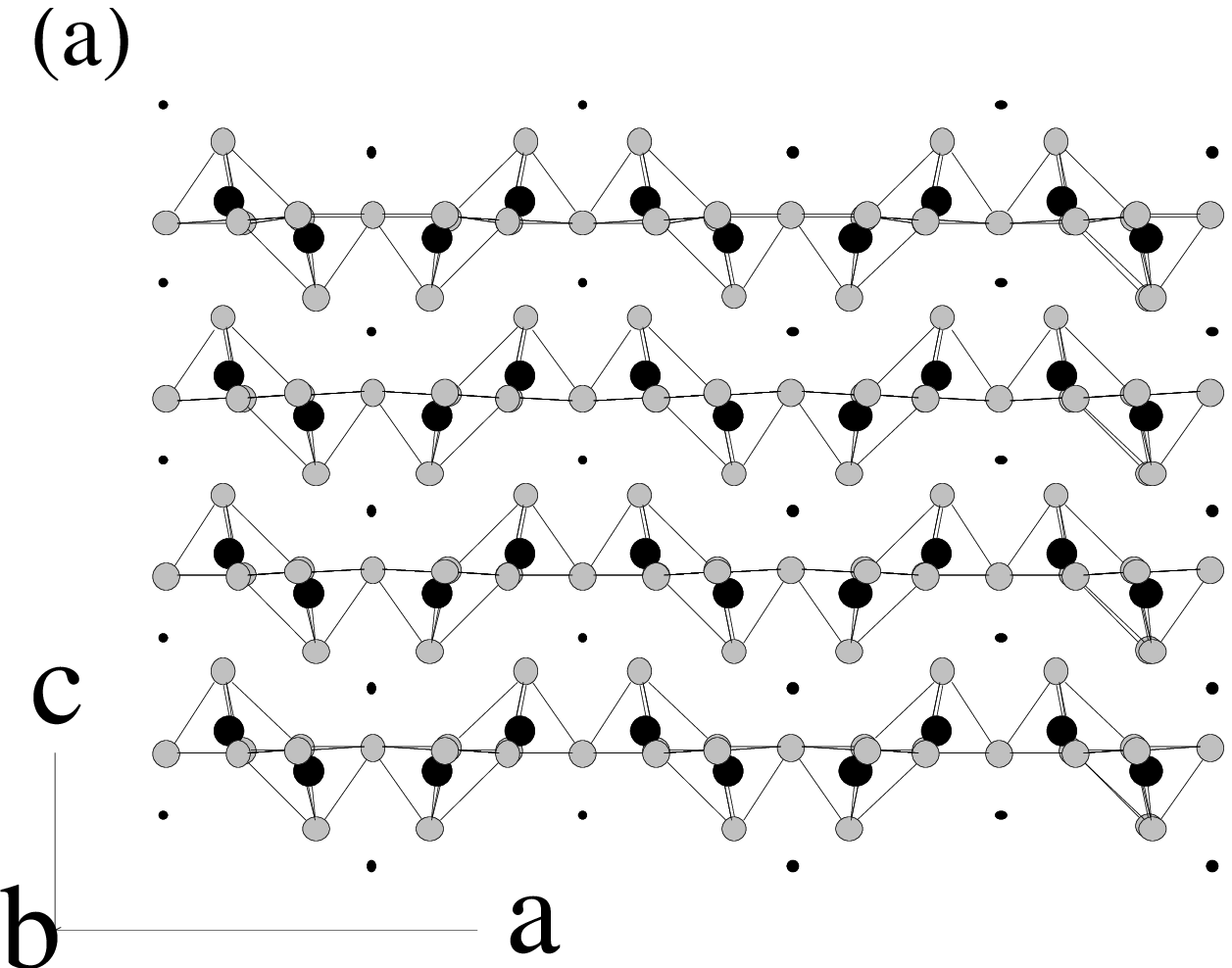}}
\resizebox{5.1cm}{3cm}{\includegraphics{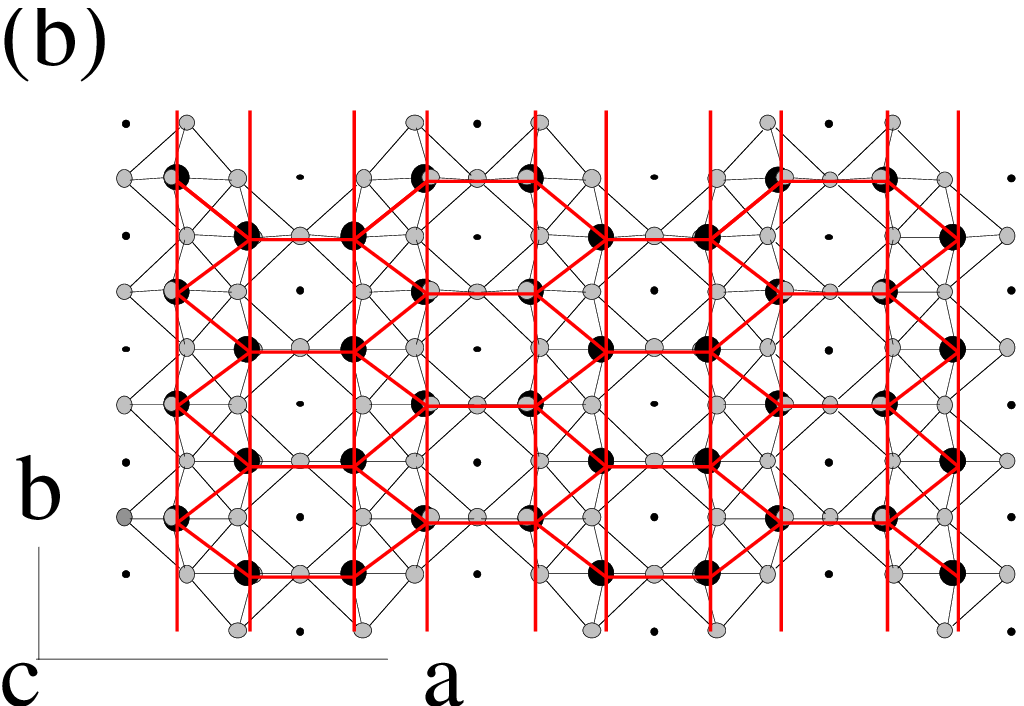}} 
}
\caption{Schematic structure of $\alpha^{'}N\!aV_2O_5$, (a) along the
(a,c) plane and (b) along the (a,c) plane. The oxygen atoms are denoted by
open circles, the vanadium atoms by filled circles and the sodium
atoms by dots.}
\label{fig:nav2o5}
\end{figure}

The $\alpha^\prime N\!aV_2O_5$ crystal is formed by layers of $VO_5$
square-pyramids stacked along the $c$ axis. The oxygen atoms of the
pyramid basis form a quasi-square planar lattice along the $a$ and $b$
directions.  The pyramids are alternatively pointing on top and below
these planes. Periodical vacancies of the $VO$ top of the pyramids are
replaced by $N\!a^+$ ions forming chains parallel to the b axis (see
fig.~\ref{fig:nav2o5}). This system can be modeled as planes of
weakly coupled ladders, the low energy physics being supported by the
$d_{xy}$ magnetic orbitals of the vanadium atoms and by the $p_y$
orbitals of the bridging oxygen atom on the ladders
rungs~\cite{vana3}.

Following a simple chemical analysis, the system was given for a long
time as a fully ionic system, with a formal charges repartition of
$N\!a^{1+}(V_2)^{9+}O_5^{2-}$ and a mixed valency on the vanadium
atom. Later on, quantum chemical calculations~\cite{vana} showed that
the vanadium-apical oxygen ($V-O_{ap}$) bond is in fact a multiple
strongly covalent bond with a small degree of polarization (about $1/2
\bar{e}$). The system should then be seen as $(V\equiv
O_{ap})^{2+\;{\rm or}\;3+}$ molecules on an oxygen $O^{2-}$ 
square lattice. A modification of the charge repartition in the
average $(VO_{ap})^{2.5+}$ group, from $V^{4.5+}O_{ap}^{2-}$ to
$V^{3+}O_{ap}^{0.5-}$ is considerable and one can expect that it
strongly affects the potential felt by the magnetic orbitals. Indeed
the electrostatic potential (computed by the Ewald summation
procedure) felt by the atoms supporting the active orbital goes from
$Pot(V)=1.735$a.u. and $Pot(O_{br})=-1.103$a.u. (where $O_{br}$ refers
to the oxygen atom bridging the two vanadium atoms on the rungs) when
formal charges are used to $Pot(V)=1.242$a.u. and
$Pot_(O_{br})=-0.930$a.u. when the polarity of the Vanadium-apical
oxygen bond has been corrected. One notices immediately that the
difference of the electrostatic potential felt by the $V$ and $O_{br}$
is increased by more than $30\%$ when formal charges are used and thus
that the energy difference between the $V d_{xy}$ orbital and the
$O_{br} p_y$ orbital should be strongly affected.

\begin{figure}[h]
\centerline{
\resizebox{4cm}{2.0cm}{\includegraphics{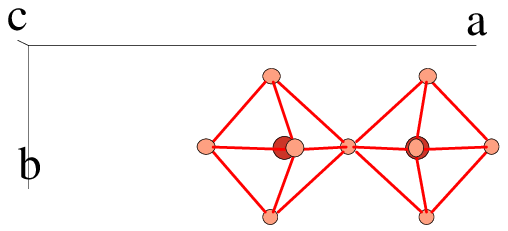} } 
\hspace*{1eM}
\resizebox{4cm}{2.0cm}{\includegraphics{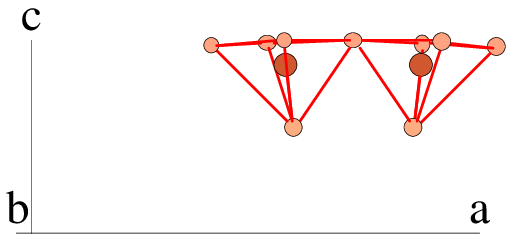}}}
\caption{Crystal fragment used for the calculations of the rung
parameters, along the (a,b) and (a,c) planes. The open circles
represent the oxygen atoms and the filled circles the vanadium atoms.}
\label{fig:rung_vana}
\end{figure}

\begin{figure}[h]
\centerline{
\resizebox{3.2cm}{7cm}{\includegraphics{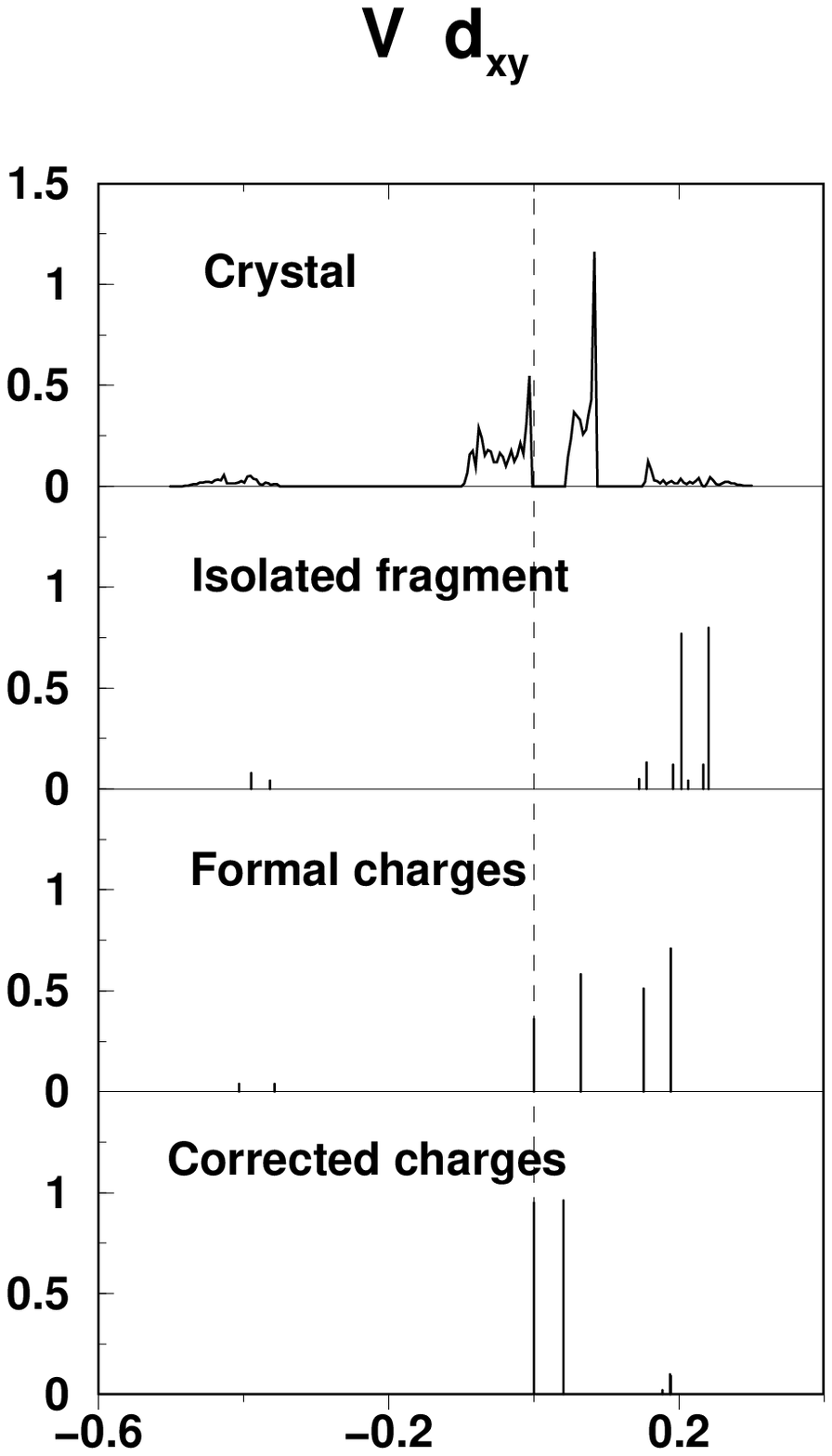}} \hspace{1eM} 
\resizebox{3.2cm}{7cm}{\includegraphics{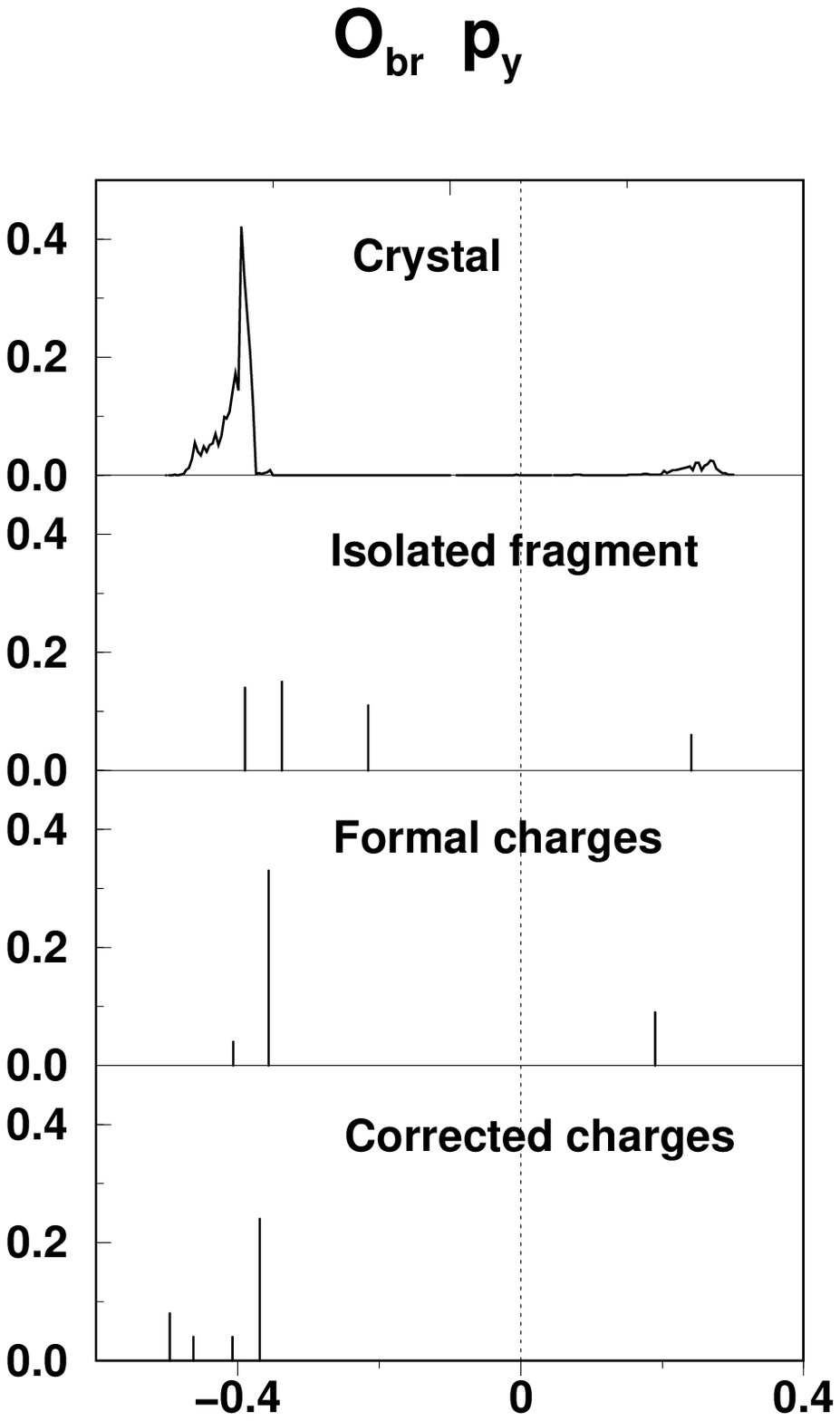} }}
\caption{Projected DOS on the magnetic atomic orbitals of the vanadium
atoms and the bridging oxygen atom in the rung geometry.  From top
to bottom~: periodic crystal HF calculation,  isolated fragment, 
embedded fragment with formal charges and embedded fragment with the
correct polarity of the $V\equiv O_{ap}$ bond. The dotted vertical
line displays the Fermi level.}
\label{fig:dens_vana}
\end{figure}

Figure~\ref{fig:dens_vana} reports the projected densities of states
on the rung active orbitals as computed at the Hartree-Fock level from
a periodic calculation, and from three fragment calculations with
different embeddings~: without embedding (isolated fragment), with an
embedding using the formal charges Madelung potential and with an
embedding using the charges corrected to take into account the
covalency of the $V\equiv O_{ap}$ bond. The rung fragment has been
chosen as including the $VOV$ rung and the set of first neighbors
atoms (see fig.~\ref{fig:rung_vana}).
One sees immediately that the isolated fragment DOS are very different
from the periodic DOS, emphasizing, if necessary, the crucial
importance of the embedding. Indeed, in the isolated fragment, even the
Fermi level orbitals ---~Highest Occupied Molecular Orbital (HOMO) and 
Lowest Unoccupied Molecular Orbital (LUMO)~--- are not the
correct ones, i.e. the bonding, anti-bonding combinations of the
vanadium $d_{xy}$ orbitals. The fragment, embedded with the formal
charges Madelung potential, yields a reasonable representation of the
occupied orbitals, the HOMO is correctly of vanadium $d_{xy}$ nature
and the bridging oxygen $p_y$ orbital is at the correct energy
difference. However, the unoccupied part of the spectrum is much more
troublesome, in particular the contribution to the virtual orbitals of
the vanadium $d_{xy}$ orbitals is totally unrealistic. We will see in
table~\ref{tb:vana_res} how this point can affect the local
spectroscopy of the system and in particular the local-singlet
local-triplet energy difference associated with the super-exchange
coupling. Finally the embedding where the polarity of the $V\equiv
O_{ap}$ bond is set to a correct value yields reasonable accuracy of
the DOS for both the occupied and unoccupied parts of the spectrum.

Table~\ref{tb:vana_res} displays the results for the local-singlet
local-triplet excitation energy ($J$) and the first local
doublet-doublet excitation energy ($2t$) of the rung fragment,
supporting respectively $4$ and $3$ magnetic electrons on $3$ magnetic
orbitals.  One sees immediately the crucial role played by the
embedding, especially on the effective magnetic exchange
integral. Indeed the singlet-triplet excitation energy is increased by
more than $70\%$ when formal charges are used.

\begin{table}[h]
\begin{tabular}{c|cc}
Charge repartition  & J (meV) & 2t (meV) \\ 
in the Madelung Potential & & \\ \hline	
$N\!a^{1+}\, (V^{4.5+})_2\, (O_{ap}^{2-})_2 \,(O^{2-})_3$ & -510 & -1162 \\ 
$N\!a^{1+}\,(V^{3+})_2\, (O_{ap}^{0.5-})^{2-})_2\, (O^{2-})_3$ & -294 & -1076
\end{tabular}
\caption{Local excitations energies on the rungs of the $\alpha^\prime
N\!aV_2O_5$ compound as a function of the embedding Madelung
potential. $J$ singlet-triplet excitation energy of the embedded
$(VO)_2^{4+}(O^{2-})_7$ fragment, $2t$ first double-doublet excitation
energy of the embedded $(VO)_2^{5+}(O^{2-})_7$ fragment. }
\label{tb:vana_res}
\end{table}

\subsection{The influence of geometrical details~: 
the incommensurate $S\!r_{14}C\!u_{24}O_{41}$}

\begin{figure}[h]
\centerline{\resizebox{8.1cm}{5.8cm}{\includegraphics{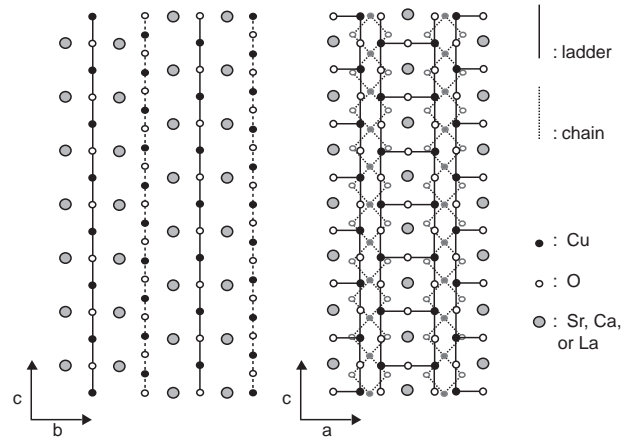}}}
\caption{Schematic structure of the incommensurate
$S\!r_{14}C\!u_{24}O_{41}$ compound along the (a,c) and (b,c)
planes. Note that the ladder and chain planes have been superposed in
the (a,c) view.}
\label{fig:srcuo1}
\end{figure}

In the light of the previous example one may wonder how much the
details of the Madelung part in the embedding can influence the
computed local excitations.  This point is of particular importance
for the incommensurate systems composed of several electronically
quasi-independent subsystems with incommensurate periodicity. Indeed,
it is of crucial importance to know up to which point the
incommensurability influences the periodicity of the local effective
interactions within one of the subsystems. In other words, what is the
amplitude of the magnetic exchange, hopping, etc. modulation in one
subsystem due to the electrostatic influence of the other subsystem.
An interesting candidates to study this problem is the super-conducing
ladder-chain copper-oxide family $S\!r_{14-x}C\!a_x C\!u_{24}O_{41}$.
Indeed, these compounds are constituted of alternate layers of
weakly-coupled doped spin-ladders and weakly-coupled doped spin-chains
(see figure~\ref{fig:srcuo1}). We will use for this study the totally
saturated $x=0$ compound in an ideal geometry where the modulation of
the chains (resp. ladders) geometry around their average structure has
been ignored. In this ideal system the two isolated subsystems are
perfectly periodical and the modulation of the local excitations
energies along the chain (resp. the ladder) can only be attributed to
the non-periodicity of the electrostatic influence of the other
subsystem.

The incommensurability of such a system does not allow us to run
benchmark periodic calculations in order to check the validity of our
embedding. However in the copper oxides, the oxidation numbers of the
different atoms ($C\!u$, $O$ as well as alkaline-earth elements) do
not seem to raise much controversy. Indeed, both the chemical analysis
and the {\em ab initio} calculations on similar systems such as the chain
system $S\!r_2C\!uO_3$~\cite{Xino1}, the ladder systems $S\!r C\!u_2
O_3$~\cite{Xino2}, or even the perovskites super-conducting parent
compounds such as the $H\!g B\!a_2 C\!u_n O_{2n+2+\delta}$
family~\cite{cuo}, yield $C\!u^{2+}$ and $O^{2-}$ oxidation states,
which we will assume for the present system.

\begin{figure}[h]
\centerline{\resizebox{6.9cm}{4.8cm}{\includegraphics{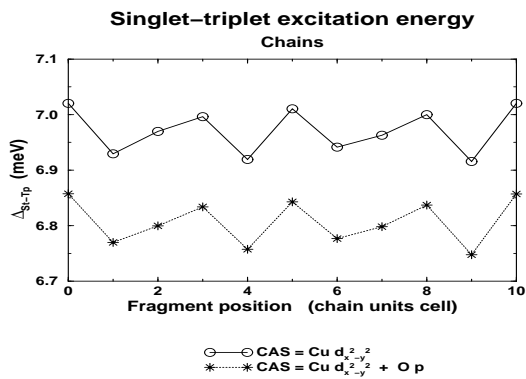}}}
\caption{Effective magnetic exchange integrals between chain nearest
neighbor copper atoms, computed at the CASSCF level. Circles~: the CAS
only contains the magnetic copper $d_{x^2-y^2}$ orbitals, stars~: the
CAS contains in addition the two bridging $p$ oxygen orbitals. The
shift between the chain and ladder unit cells reference point
corresponds to $7n/10 - Int\left[ n7/10\right]$, where $n$ is the
present graph abscissa.}
\label{fig:srcuo_ch_cas}
\end{figure}

Figure~\ref{fig:srcuo_ch_cas} shows the modulation of the local
singlet-triplet energy difference (computed at the Complete Active
Space Self Consistent Field (CASSCF) level) on the chain subsystem
fragments as a function of their reference unit cell along the chain 
subsystem. The CAS space includes either only the copper magnetic
$d_{x^2-y^2}$ orbitals (the $x$ and $y$ axes corresponding to $a$ and
$c$ crystallographic axes in figure~\ref{fig:srcuo1}) or as well the
orbitals of the bridging oxygen atoms ---~supporting the
through-bridge, super-exchange mechanism. As for the vanadium oxide
the embedded fragments have been chosen such as to include two nearest
neighbors magnetic sites (here the copper atoms), the bridging atoms
(here one oxygen atom on the ladders and two oxygen atoms on the
chains) and the first neighbor atoms to the previously cited
ones. The ladder and chain fragments are $C\!u_2O_7$ and $C\!u_2O_6$
respectively.
Unlike the change in the polarization of the $V\equiv O_{ap}$ bond in
$\alpha^\prime N\!a V_2 O_5$, the modification of the Madelung field
resulting from the relative shift between the ladders and the chains
positions induces a very small (negligible) effect on the local
excitation energies. Indeed, the maximal relative variation of the
singlet-triplet energy difference is smaller than $1.5\%$ along the
chain, and negligible along the ladders (not shown).  This point can
be easily understood by looking at the Madelung potential felt by the
different copper atoms along the chain or the ladder. Indeed, the
potential modulation is reported in table~\ref{tbl:mod}.
\begin{table}[h]
\begin{tabular}{|rr||rr|}
\multicolumn{2}{c||}{Chains} & \multicolumn{2}{c}{Ladder rungs} \\ 
Position (a.u.) & Potential (a.u.) & Position (a.u.) & Potential (a.u.) \\[1ex]
\hline
   0.7215601 &     1.9155856 &   1.8537347 &     1.81953871  \\
   5.9126401 &     1.9155567 &   9.2697116 &     1.81952754  \\
  11.1037201 &     1.9156425 &  16.6856884 &     1.81952495  \\
  16.2948001 &     1.9155681 &  24.1016653 &     1.81954098  \\
  21.4858801 &     1.9155959 &  31.5171231 &     1.81951733  \\
  26.6769601 &     1.9155856 &  38.9331000 &     1.81952982  \\
  31.8680401 &     1.9155567 &  46.3490769 &     1.81951839  \\
  37.0591201 &     1.9156425  \\
  42.2502001 &     1.9155681  \\
  47.4412801 &     1.9155959  
\end{tabular}
\caption{Electrostatic potential in atomic units for the copper atoms
along a chain and one side of a ladder. The incommensurate structure
have been approached by the usual $10:7$ commensurate one in the
chain/ladder direction. The $10$ chain cells, $7$ ladder cells
mismatch is of $0.145$\AA~out of $27.45$\AA. }
\label{tbl:mod}
\end{table}
One sees immediately, that the Madelung potential variation induced by
the other subsystem is very small, being of the order of
$10^{-4}$a.u. along the chains and of $10^{-5}$a.u. along the
ladders. However, it should be noticed that the relative variation of
the spectroscopic parameters, such as the effective magnetic exchange,
is two orders of magnitude larger than the relative variation of the
Madelung potential, pointing out the extreme sensitivity of
these local excitation energies to the electrostatic potential.

\subsection{The influence of anions TIPs~: the $P\!t_2I(dta)_4$}

Finally we will analyze the importance of the short-range effects, as
modeled by the TIPs, on local excitations.  These effects, excluding
the fragment electrons from the space  supposed to be occupied
by the electrons of the rest of the crystal, are much more
important for cations than for anions. Indeed, while the positive
charges of the former have a tendency to attract the fragment
electrons, the negative charges of the latter already repulse
them. Therefore, we will not discuss the cations case for which it is
commonly admitted that these short range exclusion effects are
crucial, and we will rather analyze the effects due to anions, more
subject to controversy.

\begin{figure}[h]
\centerline{\resizebox{6cm}{6cm}{\includegraphics{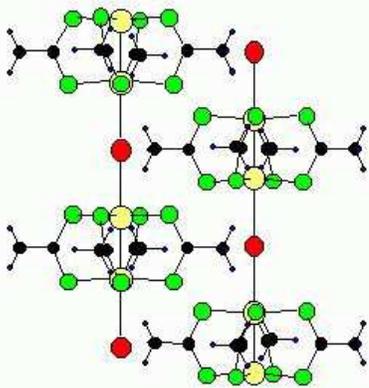}}}
\caption{$P\!t_2I(dta)_4$ crystal structure. The large empty circles
are the $P\!t$ atoms, the large dark gray circles are the $I$ atoms,
the middle size light gray atoms are the $S$ atoms, the small black
circles are the $C$ atoms and the tiny black circles are the $H$
atoms.}
\label{fig:pt2i}
\end{figure}

For this purpose we have chosen the $P\!t_2I(dta)_4$
compound~\cite{dta} since the influence of the large $I^-$ ion, first
neighbor of the $P\!t^{3+}$ magnetic centers, can be expected to be
rather important.  The $P\!t_2I(dta)_4$ system is a
quasi-unidimensional system built from the alternation of platinum dimers
and iodine anions. The coordination spheres of the $P\!t$ atoms are
completed by four $\mu$-bridging dithioacetate ($dta$) ligands (see
figure~\ref{fig:pt2i}).  The iodine anion is located at the mid center
position between two $P\!t_2$ dimers. Its distance to the closest
platinum atom is only $2.98$\AA, while the sum of its ionic radius
($2.16$\AA) and the ionic radius of the $P\!t^{2+}$ cation ($0.98$\AA)
is $3.14$\AA.
The computed fragment is composed of two $P\!t$ atoms bridged by the
four $dta$ ligands. In this system the electronic structure is
supposed to be essentially supported by the $d_{z^2-{1\over
2}\left(x^2+y^2\right)}$ orbitals of the $P\!t$ atoms (the
quantification axis being set to be the chain axis). The local singlet
to triplet excitation energy of the $\left[P\!t_2 (dta)_4\right]^{2+}$
system has been computed with and without $I^{-}$ TIP~\cite{imoins} on
the first iodine neighbors on both sides of the fragment. The system
has been found to have a metal-ligand triplet ground state. It is
noticeable that the singlet, first excited state is also a
metal-ligand strongly open-shell state. The triplet to singlet
excitation energy, computed at the CASSCF level (the CAS being defined
by $4$ electrons in $3$ orbitals) has been found to be of $46$meV when
the $I^-$ anions are only represented by a $1/R$ repulsive potential,
that is by punctual charges. When $I^-$ TIPs are used, it becomes
$50$meV. As can be expected, the TIPs destabilize more the singlet
state than the triplet state. Indeed, while the two states have the
same average number of electrons, the larger charge fluctuations in
the singlet state increases the probability of the configurations
having an extra electron on one of the $P\!t$ sites, yielding larger
repulsive contributions due to the $I^-$ TIPs.  It can therefore be
expected that on systems where the singlet state is less open-shell
than this one, the first neighbor anions exclusion effects are larger
than in the present case, reducing the singlet to triplet excitation
energy when the coupling is anti-ferromagnetic and increasing it for
ferromagnetic couplings. However, the absolute value of the TIPs
effect ($4$meV) is quite small and can be expected never to be very
large (since in the present case the iodine-platinum distance is
already smaller than the sum of the respective ionic radii). One can
therefore conclude, that the cations TIPs will be of importance
essentially when the excitation energies sought at are small (of the
order of the tens of milli-electron-Volts or smaller).

\section{Conclusion}

We have studied the effects of the environment on the local
spectroscopy of strongly correlated extended systems, as treated in
embedded fragment methods. A particular interest have been devoted to
strongly correlated systems and the local effective interactions
(magnetic exchange, hopping, repulsion, etc) between the {\em active}
electrons responsible for the low energy physical properties. We have
analyzed the different contributions of the rest of the system on the
considered fragment, originating both from the average electrostatic
and spin fields and from theirs fluctuations. A multi-polar analysis
showed that the dominant terms of fields fluctuations contributions
scales as a charge-dipole contribution ($1/R^4$). These terms decrease
rapidly as a function of the fragment response excitation distance $R$
and can therefore be neglected except from the short range
contributions. On the contrary, the effects of the average fields are
quite strong and their long range contributions, originating in the
Madelung field decrease very slowly. These effects (both the short
ranged quantum orthogonality and quantum exchange and the long range
Madelung potential) can however be treated at the mono-electronic
level and thus act essentially on the orbitals shapes and energies. We
have therefore defined a discrete function, equivalent for the finite
systems to the Density of States in periodic materials, and based a
criterion for testing the embedding quality on the concordance of the
infinite system DOS and its discrete equivalent on the embedded
fragment. A detailed analysis of different systems have shown that
unlike what is commonly assumed in the community, the definition of
the Madelung potential by the formal charges issued from simple
chemical considerations is not accurate. Indeed, these charges usually
overestimate the charge transfers, often miss the possibility of
partial covalent bonds, and in any cases are unable to evaluate the
degree of polarization of the latter. From the study of the $N\!a
V_2O_5$ compound, we have shown that the crucial importance of these
problems.  Indeed, the orbital shape and energy (and in consequence to
the discrete density of states) are very sensitive to these charge
transfer problems, as well as the local excitation
energies. Let us point out that the singlet to triplet local excitation
energy on the $N\!a V_2O_5$ rungs differ from a factor $1.7$ when the
Madelung potential is defined from formal charges and when the partial
covalency of the vanadium apical-oxygen bond is correctly defined.  On
the contrary, other effects such as the short range exclusion effects
due to the fragment first neighbor anions are very small and dominated
again by the $1/R$ electrostatic repulsion term.

In conclusion we would like to emphasize again the importance of a
correct evaluation of the charge transfer processes and a validation
of the embedding by checking the embedded fragment orbital shapes and
energies against the infinite system density of states.

\acknowledgments We thank Dr. D. Maynau for providing us with the
CASDI set of programs used in some of the presented calculations.

\end{document}